\documentclass{emulateapj}
\usepackage{apjfonts}

\newcommand{\mc}{\multicolumn}

\newcommand{\myemail}{chris.willott@nrc.ca}
\def\chandra{{\it Chandra~}}

\def\co21{CO\,(2-1)}

\shorttitle{No evidence of black holes in most z=6 galaxies}
\shortauthors{Willott}

\begin{document}


\title{No evidence of obscured, accreting black holes in most z=6 star-forming galaxies}


\author{
Chris J. Willott\altaffilmark{1},
}

\altaffiltext{1}{Herzberg Institute of Astrophysics, National Research Council, 5071 West Saanich Rd, Victoria, BC V9E 2E7, Canada; \myemail}

\begin{abstract}

It has been claimed that there is a large population of obscured,
accreting black holes at high-redshift and that the integrated black
hole density at $z=6$ as inferred from X-ray observations is
$\sim100$ times greater than inferred from optical quasars. I have
performed a stacking analysis of very deep \chandra X-ray data at the
positions of photometrically-selected $z=6$ galaxy candidates. It is
found that there is no evidence for a stacked X-ray signal in either
the soft (0.5-2 keV) or hard (2-8 keV) X-ray bands. Previous work
which reported a significant signal is affected by an incorrect method
of background subtraction which underestimates the true background
within the target aperture. The puzzle remains of why the $z=6$ black
hole mass function has such a flat slope and a low normalization
compared to the stellar mass function.
\end{abstract}


\section{Introduction}

The intimate relationship between galaxies and their nuclear black
holes is a key, but as yet unsolved, issue in astrophysics. Whilst
there is a tight correlation between stellar velocity dispersion and
black hole mass in the local universe (Ferrarese \& Merritt 2000;
Gebhardt et al. 2000), it is still unknown how this relationship
evolves at higher redshifts when most star formation, galaxy assembly
and black hole accretion occurred. Determining the evolution will
constrain which of the many theoretical explanations for the
relationship are responsible.

Willott et al. (2010) showed that the $z=6$ black hole mass function
based on optical quasars is substantially lower than expected based on
the $z=6$ stellar mass function (Stark et al. 2009) if the local
correlation between black hole and galaxy mass does not evolve at
high-redshift. On the other hand, the masses of black holes in
luminous $z=6$ quasars appear to be higher than locally for a given
galaxy mass (Wang et al. 2010), although these measurements are
potentially influenced by the selection bias discussed by Lauer et
al. (2007). These observations could be explained by a steeper
high-redshift slope to the black hole -- galaxy mass relation or a
mass-dependent active black hole duty cycle (Volonteri \& Stark 2011;
Fiore et al. 2011a).

Fiore et al. (2011a) searched for X-ray emission at the locations of
all galaxies with photometric redshifts $z>5.8$ in the Chandra Deep
Field South (CDF-S). They found significant X-ray emission in only two
galaxies of which at least one has a plausible lower $z$ photometric
solution, enabling them to place an upper limit on the space
density of low-luminosity active galaxies at $z=6$. Treister et
al. (2011; hereafter T11) found a significant signal in the stacked
X-ray flux at the locations of 197 $z\approx 6$ Lyman break galaxies
(LBG). The data in each X-ray band was stacked
independently and they found a $5\sigma$ detection in the soft
($0.5-2$ keV) band and a $6.8\sigma$ detection in the hard ($2-8$ keV)
band. This corresponds to typical X-ray luminosities of $z=6$ galaxies
of $9.2 \times 10^{41}$\, erg\,s$^{-1}$ in the soft band and $8.4
\times 10^{42}$\, erg\,s$^{-1}$ in the hard band. The high ratio of
these luminosities indicates a hard X-ray spectrum which, considering
the sizeable $k$-correction at $z\approx 6$ (soft band is rest-frame
$3.5-14$ keV and hard band is rest-frame $14-56$ keV), implies a typical
obscuring column of $N_H > 1.6 \times 10^{24}$\,cm$^{-2}$.

This X-ray emission from $z=6$ active galactic nuclei can be used to
infer an integrated black hole mass density at that epoch, making some
assumptions about the population evolution, bolometric correction and
radiative efficiency (see T11). This results in a factor of $\approx
100$ more black hole growth at high-redshift than had previously been
inferred from the optical quasar luminosity function. The inference is
that the black hole mass function is steeper than previously thought
and dominated by low-mass, heavily obscured black holes. This steeper
function would be more in line with theoretical expectations
(Volonteri \& Stark 2011). However, the ratio of obscured to
unobscured black holes would be more than an order of magnitude larger
at $z=6$ than the values of $2-4$ observed at $2<z<5$ (Treister et
al. 2009; Fiore et al. 2011a).

None of the 197 $z=6$ galaxies were individually detected at high
significance in either band of the X-ray data (Alexander et al. 2003;
Xue et al. 2011). The relatively high significance of the stacked
signal combined with the number of objects being stacked indicates a
significant contribution from many of the galaxies (T11 estimate
$>30$\%) or else some of them would have been detected individually.
With none detected individually, but a great many lurking just below
the flux detection threshold, it implies a very steep flux
(luminosity) distribution which is surprising given the large range of
galaxy luminosity and the expected range in black hole mass, accretion
rate relative to Eddington, obscuring column, etc.

In order to better understand the origin of the stacked X-ray signal
and implications for black hole mass evolution, I have analyzed the
X-ray properties of photometrically-selected $z=6$ LBGs. In this {\it
  Letter} I describe my analysis and relate it to previous work. After
submission of this {\it Letter}, Fiore et al. (2011b) published an
independent analysis of the stacked signal from $z=6$ LBGs in the
CDF-S. Those results will be discussed in Section 3.


\section{Stacking method}

Many of the details of the method follow closely those described in
the main paper and supplementary information of T11 and the interested
reader is referred there for details. Two slightly
different methods are used, one to produce the image stack for display
purposes and one to optimally determine the stacked signal and its
noise, hereafter the S/N stack. All the stacking is performed using
the mean, rather than the median that is usually considered the more
robust estimator (White et al. 2007), due to the fact we are dealing
with low integer counts.

The list of $z=6$ LBGs used by T11 contains 151 galaxies in the CDF-S
and 46 in the Chandra Deep Field North (CDF-N). Since the CDF-S 4\,Ms
\chandra data is deeper than the CDF-N 2\,Ms data, it contributes most
of the weight and only the CDF-S field is considered here. The CDF-S
4\,Ms \chandra images and catalogs (Xue et al. 2011) were downloaded
from the public website\footnotemark.  The sample of 151 CDF-S
galaxies used was drawn from a larger pool of 355 unique CDF-S
galaxies in Bouwens et al. (2006). As in T11, galaxies are not
included if they have \chandra off-axis angles $>9$ arcmin or they
have a neighboring X-ray source within 22 arcsec in the 2\,Ms \chandra
catalog of Luo et al. (2008).

\footnotetext{http://www2.astro.psu.edu/$\sim$niel/cdfs/cdfs-chandra.html}

Due to the large variation in the \chandra point spread function (PSF)
as a function of off-axis angle, an optimal photometric aperture
extraction radius, $r$, is determined for each object position. Values
of $r$ range from 1 arcsec close to the aim-point to 7 arcsec at
$\approx 9$ arcmin off-axis. The background per pixel, determined
as described below, is subtracted. The effective exposure varies somewhat
across the field-of-view, so the net counts per pixel were divided by the
exposure time to get a count rate per Ms. Because the targets with
large off-axis angles have larger photometry apertures, they will
contain a greater background signal and hence have higher
noise. Therefore each target has a weighting factor, $w_i$, determined
by the inverse noise squared, $n_i^{-2}$. Weighting factors are used
for image stacks and S/N stacks. The total exposure time for the 151
sources is $5.4 \times 10^8$\,s, equivalent to $\approx 17$ years.

For the image stacks the weighted, background-subtracted count rate
images are averaged. Images are kept in the native 0.492 arcsec
pixels. There is no stretch of the images to correct for the varying
PSF. For the S/N stacks the signal and noise for each target are
calculated individually. The average stacked signal, $S$, is
determined from the individual target signals, $s_i$, by $S=\sum_i w_i
s_i / \sum_i w_i$ and the related noise is $N=(\sum_i w_i^2
n_i^2)^{0.5} / \sum_i w_i$ (see appendix A of T11).

The final detail of the method and the one that is most important to
this analysis is the estimation of the background count rate. The
\chandra ACIS background is extremely low and even in the 4\,Ms data,
most pixels have zero counts in each of the soft and hard bands (Xue
et al. 2011). Nevertheless, the background is the largest source of
uncertainty in the stacking analysis due to the very low signal. The
background is a mixture of instrumental effects, diffuse
galactic/extragalactic emission and unresolved extragalactic point
sources (Markevitch et al. 2003).
 
Background counts are determined from the pixel values measured within
an annulus from $2r$ to 22 arcsec around each target position. Three
methods of background subtraction are considered.

\renewcommand{\labelenumi}{\alph{enumi}}
\begin{enumerate}
\item -- The background is determined from the mean value, $\mu$, of pixels within the background annulus.
\item -- The standard deviation, $\sigma$, and mean of all pixels within a radius of 22 arcsec around each target position is calculated. Any pixel within this circle with a value greater than $\mu+3\,\sigma$ has its value set to $\mu$. For the rare cases where $\mu+3\,\sigma <1$, only pixels with 2 or more counts were clipped. This method suppresses noise in the background annulus, but will also remove real signal from the stacked position.
\item -- The standard deviation and mean of all pixels within the background annulus from $2r$ to 22 arcsec is calculated. Pixels in the background annulus with value greater than $\mu+3\,\sigma$ had their values set to $\mu$. Again, for rare cases where $\mu+3\,\sigma <1$, only pixels with 2 or more counts were clipped. This is the background method adopted by T11. This method has the seemingly attractive behaviour of eliminating noise and undetected point sources in the background determination whilst not eliminating flux from the target position. However, it will be shown that this leads to a considerable bias in a stacking analysis. 
 
\end{enumerate}

\begin{figure}
\vspace{0.2cm}
\resizebox{0.48\textwidth}{!}{\includegraphics{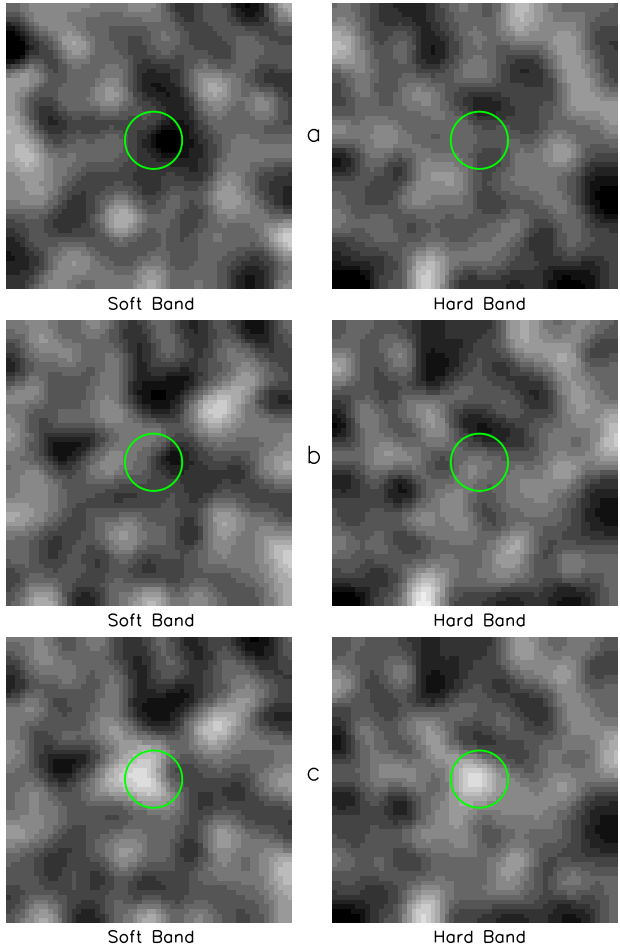}}
\caption{Weighted, stacked CDF-S 4\,Ms data at the locations of the
  151 $z=6$ galaxies in the sample of T11 smoothed by a 7 pixel
  (3.5 arcsec) Gaussian.  Each image is 30 arcsec on a side and the
  green circle has radius 3 arcsec. The greyscale range is from
  median\,$-3\,\sigma$ (black) to median\,$+5\sigma$ (white). The left
  panels are for the soft band ($0.5-2$ keV) and the right panels are
  for the hard band ($2-8$ keV). The upper two panels show method (a)
  background subtraction, i.e. no $3\,\sigma$ clipping. The middle two
  panels show method (b) where pixels within a radius of 22 arcsec of
  the target position have all pixels $>3\,\sigma$ from the mean set to
  the mean value. The lower two panels show method (c) where only pixels
  in the background annulus (from 2$r$ to 22 arcsec) are subject to
  $3\,\sigma$ clipping. It is clear that in both bands, there is only
  evidence for a significant central stacked signal when only the
  background annulus is subject to 3$\sigma$ clipping.
\label{fig:stackim}
}
\end{figure}

\section{Results}

Image stacks using the three different background determination methods
are shown in Fig.\,\ref{fig:stackim}. Each image displays a 30 arcsec
region and has been smoothed by a 7 pixel Gaussian. Methods (a) and
(b) show no sign of an excess of flux at the center for either
band. Method (c) shows significant positive signal in both bands.

Values for the stacked S/N for each method and band are given in
Table\,1. These confirm that the only significant signal is for method
(c). Indeed, the soft band for method (a) shows a negative flux,
although it is only significant at S/N$=-1.3$ and hence not
particularly unlikely to occur by chance. The values of signal and
background for method (c) have been compared to those for the same
targets in T11 and are very similar (Treister, priv. comm.). The
stacked S/N for method (c) of 4.7 for the soft band and 6.1 for the hard
band are comparable, but slightly lower, than the results for the full
sample (including CDF-N) of 197 sources which have S/N of 5 for soft
and 6.8 for hard in T11.

There is an obvious reason why the stacked signal only appears for
method (c). In this case, the background annulus has had $>3\,\sigma$
pixels clipped, but the same clipping has not been applied to the
target aperture. Whilst for each target this only makes a small
difference to the background level ($\approx 10$\%), it is a
systematic effect that builds up in the stack. $\sigma$ clipping makes
sense for some data with sharp noise spikes, e.g. optical data
affected by cosmic rays, but if it is to be applied then it must also
be applied to the target positions to avoid a cumulative positive bias
(as in method b), and it must be known that this will not remove
signal from the sources being stacked. In the CDF-S \chandra case,
where much of the background is due to faint sources below the
detection threshold, it makes more sense to do no $\sigma$ clipping at
all, i.e. method (a), and to leave the faint sources in both the
target and background data, knowing that on average they will cancel
out.  Therefore, the optimum results are those of method (a) with
measured S/N of $-1.3$ and 0.1 in the soft and hard bands,
respectively. These correspond to $3\sigma$ limits on the count rate
per galaxy of $<2.4 \times 10^{-7}$\,ct\,s$^{-1}$ and $<4.2 \times
10^{-7}$\,ct\,s$^{-1}$ in the soft and hard bands, respectively, where
a correction has been applied for counts outside the photometric
apertures. This hard band $3\sigma$ upper limit is only half the
reported hard band signal of T11.

Fiore et al. (2011b) also performed stacking of the CDF-S \chandra
data at the known positions of $z=6$ LBGs from Bouwens et
al. (2006). They used method (a) stacking without any optimization for
the off-axis dependent PSF. They considered two samples with different
constraints on how close the LBGs can lie to a neighboring X-ray
source in Xue et al. (2011). Their samples contained 210 (77) galaxies
with no neighbor closer than 10 (22) arcsec. The stacked images for
each sample in a range of X-ray bands show no significant positive
signal at the center. Fiore et al. (2011b) quote $3\sigma$ limits for
the average count rate from the 210 galaxies of $<3.4 \times
10^{-7}$\,ct\,s$^{-1}$ and $<5.8 \times 10^{-7}$\,ct\,s$^{-1}$ in the
soft and hard bands, respectively. The results presented here are
consistent with, and a little more stringent than, those of Fiore et
al., mainly due to the use of optimized apertures and weighting.

\begin{table}
\caption{Statistics of \chandra CDF-S $z=6$ galaxy stacking}
\vspace{-0.5cm}
\begin{center}
\begin{tabular}{ccc}
\colrule\colrule
\mc{1}{c}{Background} &\mc{1}{c}{Soft band} &\mc{1}{c}{Hard band} \\
\mc{1}{c}{method}     &\mc{1}{c}{S/N}           &\mc{1}{c}{S/N}    \\
\colrule
(a) & $-1.3$ & $0.1$  \\
(b) & $-0.1$ & $1.1$  \\
(c) & $4.7$  & $6.1$    \\
(c*) & $4.7$  & $6.7$    \\
\colrule
\end{tabular}
\end{center}
{\sc Notes.}--- Measured S/N of the stacked signal of the 151 $z=6$ galaxy positions in T11 for the soft and hard bands using the three methods of background subtraction. (c*) gives the results for the 141 $z=6$ galaxy positions which are further than 19 arcsec from an X-ray source in the 4\,Ms catalog of Xue et al. (2011).
\label{tab:param}
\end{table}

It is worth noting that the distribution of the S/N of each target as a
function of the weighting factor is a useful diagnostic for a true
versus spurious signal. I have applied method (a) stacking to the
samples of infrared star-formation rate excess (ISX) and infrared
star-formation rate normal (ISN) galaxies in the CDF-S stacked by Luo
et al. (2011). Both samples show significant stacked signals because
these are galaxies similar to those detected in the CDF-S, but with
fluxes just below the detection threshold. These samples show a positive
correlation between S/N and weighting factor, because the sources have
similar signal but those with lower weight have higher noise. In
contrast, the S/N against weighting factor using method (c)
stacking of the $z=6$ LBG sample shows a negative correlation. This is
because the lowest weight objects have larger photometry apertures $r$
and hence higher integrated backgrounds and a greater problem of
insufficient background removal.

T11 required the criterion that the $z=6$ galaxies not have a
neighboring X-ray source from the 2\,Ms catalog of Luo et al. (2008)
within 22 arcsec. Since the 4\,Ms catalog of Xue et al. (2011) is now
available, its use to avoid nearby X-ray sources is preferable because
it contains more sources. However, using a criterion of 22 arcsec with
the 4\,Ms catalog reduces the number of $z=6$ galaxies to stack to
110. For the following analysis, the 22 arcsec criterion is relaxed to
19 arcsec so a similar sample size of 141 LBGs is retained. The outer
limit of the background aperture is set to 18 arcsec to limit the
influence of known sources just outside the aperture. As shown in the
lowest row of Table 1, this minor change in sample and background
determination does not significantly change the results compared to
using method (c) with the 151 sources from T11.

A simple test of the robustness of any stacking method is to stack at
random positions in the sky or at positions of sources known to not
emit at the relevant waveband (e.g. White et al. 2007). The expected
signal should be zero with a distribution about zero that gives the
dispersion due to noise from the background.  To quantify the size of
the S/N bias using method (c), a Monte-Carlo process of stacking the
CDF-S \chandra maps at random locations is performed. To obtain
a similar sample size and off-axis angle distribution to T11, the
random locations are chosen to be offset from known $z=6$ galaxies
(Bouwens et al. 2006) by distances randomly drawn in the range 15 to
30 arcsec in right ascension and declination. The same selection
criteria of off-axis angle less than 9 arcmin and no neighbor X-ray
source within 19 arcsec are applied. At the new random positions the
same process of S/N stacking was carried out using method (c)
background subtraction. 

\begin{figure}
\resizebox{0.48\textwidth}{!}{\includegraphics{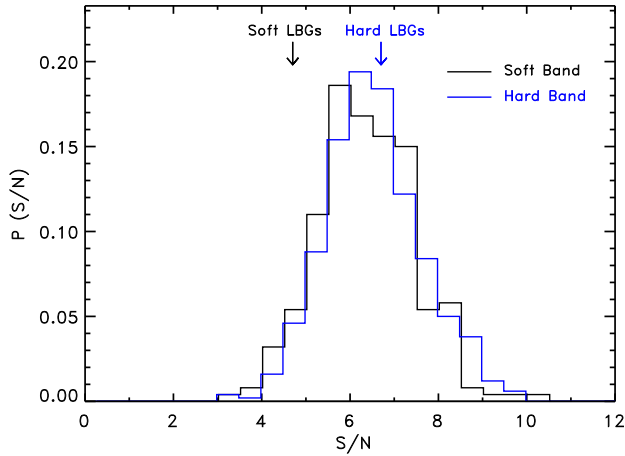}}
\caption{Probability distribution of measured S/N for stacks of $\approx 144$ random locations within the CDF-S using method (c) of $3\,\sigma$ clipping only the background annulus pixels. The soft band results are shown in black and the hard band in blue. The measured S/N of the stacks in each band at the true locations of the $z=6$ galaxies are shown by arrows. One can see that these are indistinguishable from the typical results obtained at random locations.
\label{fig:mcstack}
}
\end{figure}

This process was repeated using 500 sets of random locations and the
resulting distribution of S/N values are shown in
Fig.\,\ref{fig:mcstack}. The typical number of positions stacked was
$144\pm8$. In every single one of the 500 trials there is a
significant ($>3\,\sigma$) positive signal at random locations. The
typical values for both the soft and hard bands are S/N=6.5. For
comparison, the values found at the $z=6$ galaxy positions (c* in
Table 1) are shown by arrows. For the hard band this is at
approximately the peak of the distribution. For the soft band the
observed S/N of 4.7 is at the lowest 10\% of the distribution, so as
before it is found that these target positions just happen to be
located at mild flux deficits in the soft band images. It is concluded
that method (c) background subtraction always leads to a false
positive signal, due to under-estimation of the true background level.

\section{Conclusions}

It has been demonstrated that there is no evidence for a stacked X-ray
signal in either the soft or hard bands at the positions of
photometrically-selected $z=6$ galaxy candidates. The positive signal
determined previously is due to an under-estimation of the background
due to $3\,\sigma$ clipping of the background pixels. Stacking random
locations within the X-ray images with this method leads to similar
positive signals as found at the LBG positions. The hardness of the
X-ray signal found by T11 is a combination of the fact that the $z=6$
galaxies happen to lie at locations with below average flux in the
soft X-ray image and that the X-ray background is higher in the hard
band than the soft band, with the latter effect dominant.

The conclusion is that there is no evidence yet at $z=6$ for either a
very high black hole mass density or a sharp increase in the ratio of
obscured to unobscured black holes. The determination of the $z=6$
black hole mass function by Willott et al. (2010) based on moderate
and high luminosity optical quasars remains the most reliable estimate
of this function, despite the considerable uncertainties due to quasar
duty cycle and ratio of obscured to unobscured quasars. As noted by
Willott et al. (2010) the $z=6$ black hole mass density is
substantially lower than the $z=6$ stellar mass density expectation if
the black hole -- galaxy correlation does not evolve. The function is
also flatter than expected based on models of Eddington-limited black
hole growth. Volonteri \& Stark (2011) show that negative evolution of
the black hole -- galaxy mass correlation (or equivalently, a
substantial fraction of high-redshift galaxies without black holes)
and a steeper correlation than found locally, perhaps due to a
correlation between accretion efficiency and halo mass, are required
to fit the observed $z=6$ black hole mass function. Fiore et
al. (2011a) suggest there is a mass-dependence to the active black hole
duty cycle (similar to as observed at low-redshift) such that at
high-redshift, most high-mass black holes are active, but most
low-mass black holes are not. This would explain the flatter {\it
  active} black hole mass function and suggest there is a large,
hidden population of high-redshift black holes, but that they are not
rapidly growing.

\acknowledgments

Thanks to Ezequiel Treister for providing the data file for the CDF-S
sources used in their analysis and for discussions regarding the
differences between our stacking methods. Thanks also to Matt Jarvis
for reading the manuscript and providing comments and to the anonymous
referee for providing useful suggestions for improvement.

\end{document}